\documentclass[aps,prl,reprint,superscriptaddress]{revtex4-2}
\usepackage{graphicx}
\usepackage{subcaption}
\usepackage{comment}
\usepackage{lipsum}
\usepackage{xcolor}
\usepackage{mathtools}
\usepackage{caption}
\usepackage{subcaption}
\usepackage{hyperref}
\usepackage[normalem]{ulem}

\begin{document}

%\preprint{}

\title{Generalized tension metrics for multiple cosmological datasets}


\author{Matías Leizerovich}
\email[]{mleize@df.uba.ar}
\homepage[]{https://matiasleize.github.io/}
%\thanks{}
%\affiliation{Universidad de Buenos Aires, Facultad de Ciencias Exactas y Naturales, Departamento de Física. Buenos Aires, Argentina.}
\affiliation{Universidad de Buenos Aires, Facultad de Ciencias Exactas y Naturales, Departamento de Física. Buenos Aires, Argentina.}
\affiliation{CONICET - Universidad de Buenos Aires, Instituto de Física de Buenos Aires (IFIBA). Buenos Aires, Argentina.}

\author{Susana J. Landau}
%\thanks{}
\affiliation{CONICET - Universidad de Buenos Aires, Instituto de Física de Buenos Aires (IFIBA). Buenos Aires, Argentina.}

\author{Claudia G. Scóccola}
%\thanks{}
\affiliation{DFI-FCFM, University of Chile, Santiago, Chile}

\date{\today}

\begin{abstract}
We introduce a novel estimator to quantify statistical tensions among multiple cosmological datasets simultaneously. This estimator generalizes the Difference-in-Means statistic, $Q_{\rm DM}$, to the multi-dataset regime. Our framework enables the detection of dominant tension directions in the shared parameter space. It further provides a geometric interpretation of the tension for the two- and three-dataset cases in two dimensions. According to this approach, the previously reported increase in tension between DESI and Planck from $1.9\sigma$ (DR1) to $2.3\sigma$(DR2) is reinterpreted as a more modest shift from $1.18\sigma^{\rm eff}$ (DR1) to $1.45\sigma^{\rm eff}$ (DR2). These new tools may also prove valuable across research fields where dataset discrepancies arise.
\end{abstract}

%\keywords{}

\maketitle

\textit{Introduction ---} 
Over the past two decades, the amount and precision of cosmological data have increased dramatically. As a result, stringent constraints have been established for the cosmological parameters, including the baryon and dark matter densities and the Hubble constant. Although the $\Lambda$CDM model provides an excellent fit to most datasets, several notable discrepancies persist. These unresolved tensions remain an open question for cosmologists and continue to motivate the exploration of alternative cosmological models. The most prominent among them is the  $H_{0}$ tension: the value of the present-day Hubble parameter inferred from CMB data and assuming $\Lambda$CDM \cite{Planckcosmo2018} is inconsistent with late-time, distance-ladder determinations based on Type Ia supernovae and Cepheids \cite{Riess2022}.  Despite extensive investigation, the origin of this discrepancy remains debated \cite{2021ApJ...919...16F,2024IAUS..376...15R,2022PhR...984....1S,Vagnozzi_2023}. Such tensions motivate the development of robust, statistically sound tools to compare parameter inferences from independent datasets.

A common way to quantify discrepancies is to compare one-dimensional marginalized posteriors. For a parameter $\theta$ estimated from datasets $A$ and $B$, the rule-of-thumb estimator is defined as~\cite{2021MNRAS.505.6179L}:
\begin{equation} \label{eq: rule_of_thumb}
    N_{\sigma} = \frac{|\mu_{A}-\mu_{B}|}{\sqrt{\sigma_{A}^{2}+\sigma_{B}^{2}}}\, ,
\end{equation}
where $\mu_{A,B}$ and $\sigma_{A,B}$ denote the means and standard deviations. While simple, this approach suffers well-known limitations: marginalization can hide tensions that appear only in higher-dimensional parameter space, and the inferred discrepancy depends on the number of jointly constrained parameters.

To overcome these issues, several global tension metrics have been proposed \cite{PhysRevD.99.043506,PhysRevD.104.043504,Leizerovich:2023qqt}, including the Difference in Means statistic $Q_{\rm DM}$, which generalizes the rule of thumb to the full parameter space, but is limited to evaluating the tension between two datasets. With the rapid growth of high-quality observational datasets, there is an increasing need for methods that assess tensions involving \textit{multiple} posterior distributions simultaneously.

In this Letter, we develop a general framework for quantifying tensions among more than two posterior distributions. Within this scheme, we introduce a new estimator that extends hypothesis-testing-based global tension metrics to the multi-dataset regime. As in the case of $Q_{\rm DM}$, evaluating the tension requires formulating an appropriate null hypothesis in the full parameter space. 

The probability-to-exceed (PTE) associated with $Q_{\rm DM}$,  yields an effective $N_\sigma$ that
reduces to the rule of thumb expression if the shared parameter space is one-dimensional. In this limiting case, the resulting $N_\sigma$ can be interpreted as a measure of the separation between two one-dimensional posterior distributions. However, when the shared parameter space has dimension greater than one, the PTE-derived $N_\sigma$ no longer 
admits a straightforward interpretation as a distance metric between multivariate  posteriors. This limitation applies equally to the $N_\sigma$ values obtained from our newly proposed estimator. Therefore, establishing a clear criterion for interpreting PTE values in higher-dimensional shared parameter spaces is  an essential step toward a robust assessment of statistical tension.

A second key contribution is a geometric interpretation of our estimator in two or more dimensions of the parameter space. We illustrate this using a test case involving three posterior distributions, and defining an effective $N_\sigma^{\rm eff}$ separation between the corresponding two-dimensional confidence regions. The same geometric picture also applies to $Q_{\rm DM}$, providing---for the first time, to our knowledge---
a clear  two-dimensional geometric  interpretation 
of this widely used statistic.

\textit{Generalization of the tension estimator ---} 
We introduce a geometric framework to quantify mutual consistency among $N$ posterior distributions defined on a $D$-dimensional shared parameter space. For each pair $k = (i,j)$  of datasets, we define the \textit{tension vector} $\vec{r}_{k}$ as
\begin{equation}
    \vec{r}_{k} = \frac{1}{\sqrt{\hat{C}_{i}+\hat{C}_{j}}} (\vec{\bar{\theta}}_i - \vec{\bar{\theta}}_j)  \,\,\,\, (i \neq j).
    \label{eq:r_k}
\end{equation}
where $\vec{\bar{\theta}}_i$  and $\hat{C}_{i}$ are the mean and covariance matrix of the posterior distribution of dataset $i$ in the shared parameter space.
 The number of ordered tension vectors, $N_p$, is twice the number of pairs of datasets: $N_p= 2\binom{N}{2}= N\cdot(N-1)$, and $\vec r_{ij}=-\vec r_{ji}$; we therefore adopt a symmetric construction to avoid ordering ambiguities.

To motivate our estimator, we use an analogy with  a many-particle system. The vectors $\vec r_k$ are treated as elements of a parameter-difference space and identified with the positions of $N_p$ particles of mass $m$, which encode the weights of each dataset pair. As uniform weights suffice, we set $m=1$. The center of mass is
\begin{equation}
    \vec{R}_{CM} = \frac{\sum_{k=1}^{N_p} m_k\vec{r_k}}{\sum_{k=1}^{N_p} m_k} = \frac{\sum_{k=1}^{N_p} \vec{r_k}}{N_p}=0\, ,
\end{equation}which vanishes by antisymmetry.
We define our new estimator of the tension $\mathcal{Q}$ as the dispersion of the $\vec{r}_k$ vectors w.r.t. $\vec{R}_{CM}$, i.e., the sum of the squared tension vectors
\begin{equation}\label{eq: Q}
    \mathcal{Q} \equiv  \frac{1}{N_p}\sum_{k=1}^{N_p}|\vec{r}_k|^2 \, .    
\end{equation}
From the later, we define the dispersion tensor as 
\begin{equation} \label{eq: Cij}
    \mathcal{C}_{ab} = \frac{1}{N_p}\sum_{k=1}^{N_p} r^{a}_{k} \cdot r^{b}_{k} \, .
\end{equation}
$\mathcal{C}_{ab}$ quantifies the dispersion of $\vec{r}_k$ with respect to the origin. The indices $(a,b)$ run over the directions of the shared parameter space $(1,..,D)$. 
The dispersion tensor is related to $\mathcal{Q}$ through
\begin{equation} \label{eq: F1_formalism}
    \mathcal{Q} = \text{Tr}(\mathcal{C})= \sum_\alpha^{D} \lambda_\alpha \, ,
\end{equation}
where \{$\lambda_\alpha$\} are the eigenvalues of the tensor $\mathcal{C}$.

Each $\vec{r}_k$ is a random variable induced by the posterior samples, reflecting the statistical nature of the underlying datasets.
To quantify the tension, we formulate the problem as a hypothesis test. 
We specify a null hypothesis ($H0$) in terms of the tension vectors introduced above and determine the distribution of the test statistic $\mathcal{Q}$ under $H0$. This yields a $p$-value that determines whether $H0$ is accepted or rejected. In our context, $H0$ encodes our operational definition of concordance between posterior distributions.

Since the random vectors $\vec{r}_k$ are distributed as $\vec{r}_k \sim \mathcal{N}(\frac{1}{\sqrt{\hat{C}_{i}+\hat{C}_{j}}}(\vec{\bar{\theta}}_i - \vec{\bar{\theta}}_j),\mathcal{I}_{D\times D})$ (see Appendix), we adopt our null hypothesis, $H0$, as 
\begin{equation} \label{eq:def-H0}
    \frac{1}{\sqrt{\hat{C}_{i}+\hat{C}_{j}}} (\vec{\bar{\theta}}_i - \vec{\bar{\theta}}_j)  \simeq \vec{0} \, \, \, \forall i,j, i\neq j
\end{equation}
which states that the mean of each $\vec{r}_k$ is negligible under $H0$. Equation~\ref{eq:def-H0} formalized our definition of tension, and coincides with the condition underlying $Q_{\rm DM} $ in the case $N_p=1$.

\begin{figure}
    \centering

    % --- Fila 1: a y c ---
    \begin{subfigure}{0.49\linewidth}
        \centering
        \includegraphics[width=\linewidth]{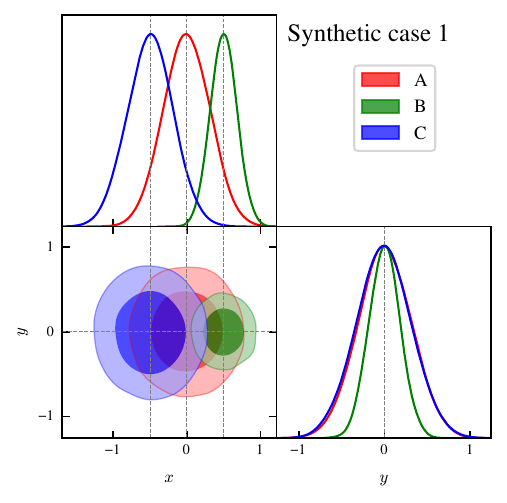}
        %\caption{}
        \label{fig:a}
    \end{subfigure}
    \hfill
    \begin{subfigure}{0.49\linewidth}
        \centering
        \includegraphics[width=\linewidth]{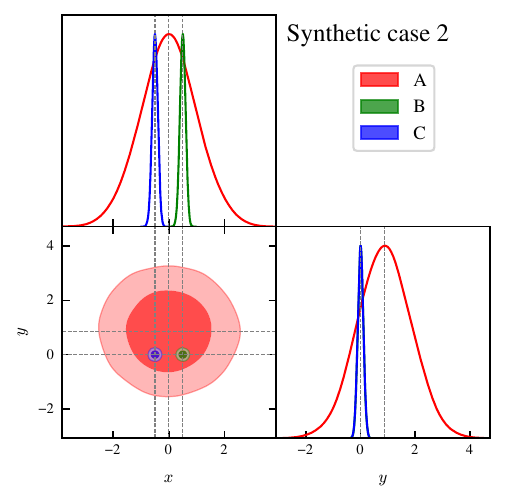}
        %\caption{}
        \label{fig:c}
    \end{subfigure}

    \vspace{0.7em}

    % --- Fila 2: b centrada ---
    \begin{subfigure}{\linewidth}
        \centering
        \includegraphics[width=\linewidth]{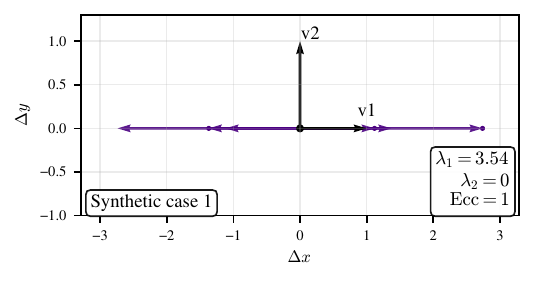}
        %\caption{}
        \label{fig:b}
    \end{subfigure}

    %\vspace{0.7em}
    %\hfill
    % --- Fila 3: d centrada ---
    \begin{subfigure}{\linewidth}
        \centering
        \includegraphics[width=\linewidth]{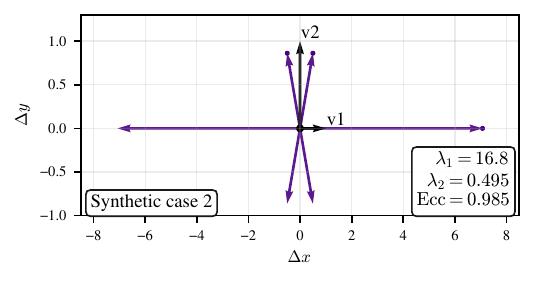}
        %\caption{}
        \label{fig:d}
    \end{subfigure}

    \caption{ Posterior distributions for three synthetic datasets in two contrasting configurations  (upper panels), and the corresponding tension vectors in parameter-difference space (lower panels). The eigenvalues of $\mathcal{C}_{ab}$ quantify both the strength and geometric structure of the multi-dataset tension.
    Eigenvectors are also shown for reference ($\vec{v}_i$).
    }
    \label{fig:4panel}
\end{figure}

In general, an analytical expression for the probability distribution function associated to $H0$ does not exist, since the random vectors $\vec{r}_i|_{H0}$ are not independent of each other. However, when all covariances are equal, the distribution under $H0$ is $\mathcal{Q}|_{H0} \sim \Gamma(D,1)$ (for more details, see Appendix), where $D$ is the dimension of the shared parameter space. Note that in the case of $N=2\, (N_p=1)$ both the statistic $\mathcal{Q}$ and its distribution under $H0$ reduce to the one of $Q_{\rm DM}$, as expected. We stress that $\mathcal{Q}$ acts as a global estimator of the tension.
Figure \ref{fig:4panel}  illustrates two representative configurations for the case of three synthetic datasets, and their mapping into the space of tension vectors.

\textit{Geometrical interpretation ---} The level of agreement or tension between inferred parameter values from different datasets is often quantified in terms of an equivalent significance $N_\sigma$, defined by analogy with a one-dimensional Gaussian variate through
\begin{equation}
1 - \mathrm{PTE} = \mathrm{Erf}\left(\frac{N_\sigma}{\sqrt{2}}\right) \, .    
\end{equation}
Here, the PTE is the probability to exceed the measured value of a given test statistic under the null hypothesis. However, care is required when comparing PTEs (and their associated $N_\sigma$ values) that arise from physically distinct situations involving different test statistics, different numbers of datasets, or different dimensionalities of the shared parameter space.
For instance, the $N_\sigma$ obtained from Eq.~\ref{eq: rule_of_thumb} is not directly comparable to that derived from the $Q_{\rm DM}$ metric, since the corresponding null hypotheses and test-statistic distributions differ if the shared parameter space is greater than 1.
A further source of confusion arises when the PTE-derived $N_\sigma$ is interpreted as a one-dimensional tension in a single parameter. 
Rather, this quantity reflects the global tension in the full shared parameter space. 
To overcome these limitations, we introduce a geometrical interpretation in which tension is represented by the distance between the 2D confidence regions of an equivalent configuration.
The interpretation applies directly to 2D shared parameter spaces and can be generalized to higher dimensions.

\begin{figure}
    \centering
    \includegraphics[width=0.7\linewidth]{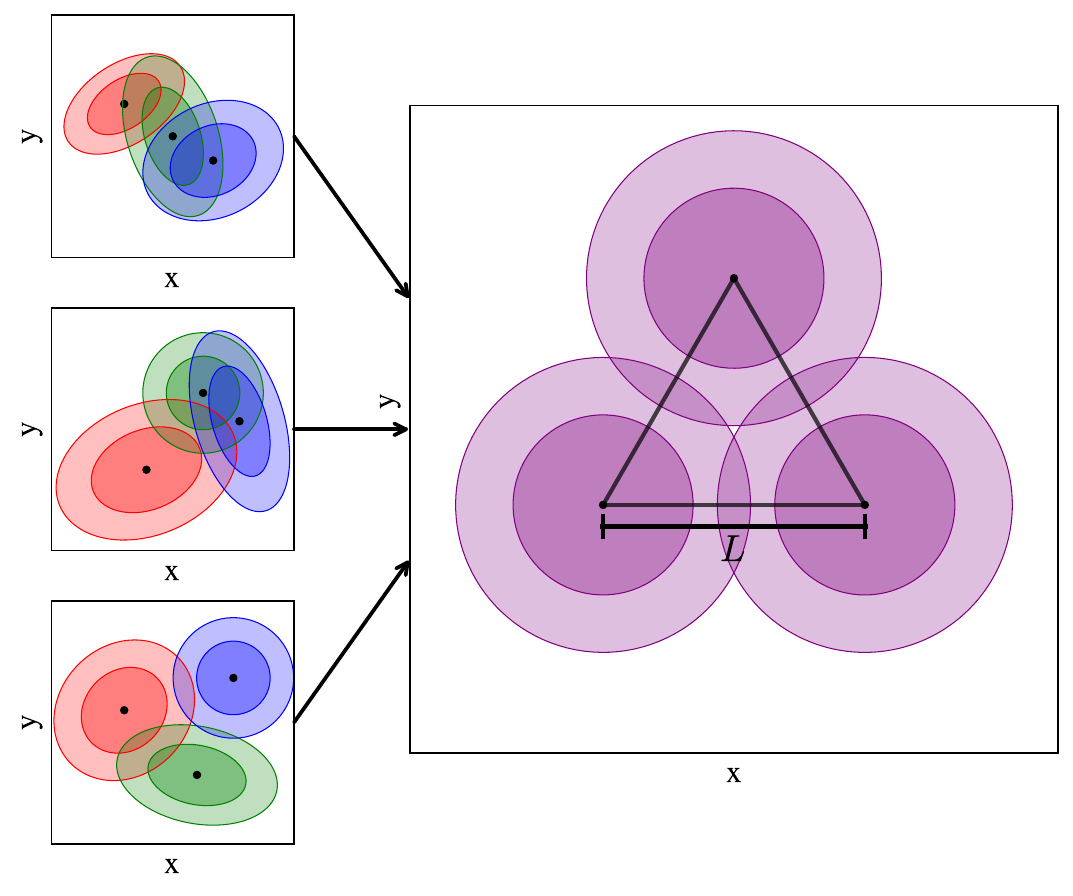}
    \caption{ Different configurations of three posterior distributions mapped to the equivalent reference setup consisting of three posteriors with identical covariances, and means positioned at the vertices of an equilateral triangle.}
    \label{fig: mapping}
\end{figure}

Our proposal is to map any arbitrary configuration of three 2D posteriors onto a reference setup consisting of three posteriors with identical covariance, chosen as $C=\tfrac{1}{2}\mathcal{I}_{D\times D}$, and with their means placed at the vertices of an equilateral triangle of side length $L$. This construction is convenient, not only because of its symmetry but also because the equality of covariances yields an analytic expression for the distribution of $\mathcal{Q}$ under $H0$. Figure \ref{fig: mapping} shows an example of such a mapping. This geometrical construction provides a simple and more accurate way to quantify the tension in the shared parameter space.

\begin{figure}
    \centering
    \includegraphics[width=\linewidth]{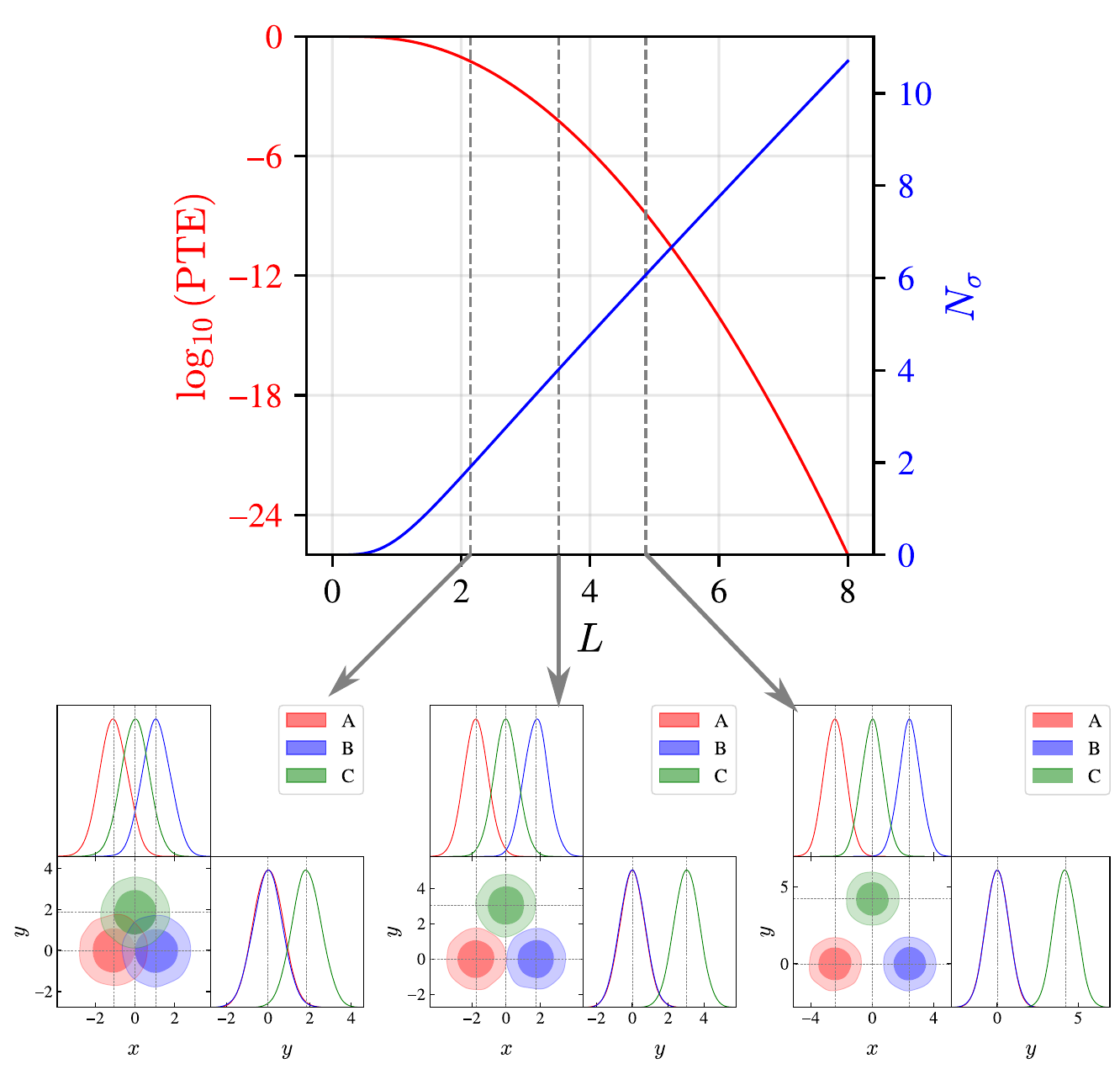}
    \caption{PTE and corresponding $N_\sigma$ as a function of the side length $L$ of the reference equilateral-triangle configuration. Dotted lines indicate values of L equivalent to $N_\sigma^{\rm eff}=\{1,2,3\}$ in the shared parameter space.}
    \label{fig:dist_between_vert}
\end{figure}

Figure \ref{fig:dist_between_vert} shows the PTE and the corresponding $N_\sigma$ as a function of the separation $L$ between the means of the reference configuration. The dotted vertical lines indicate separations that correspond to discrepancies of $1\sigma^{\rm eff}$, $2\sigma^{\rm eff}$, and $3\sigma^{\rm eff}$ significance in the shared parameter space, allowing us to define the equivalent scale
$L(N_\sigma^{\rm eff}=1)=2.143$, $L(N_\sigma^{\rm eff}=2)=3.516$, $L(N_\sigma^{\rm eff}=3)=4.864$.
Here, $N_\sigma^{\rm eff}$ denotes the effective significance at which the 2D confidence regions overlap in the two-dimensional reference configuration: $N_\sigma^{\rm eff}=1$ marks the first intersection of the $68\%$ contours, while $N_\sigma^{\rm eff}=2$ corresponds to the onset of overlap of the $95\%$ regions, and analogously for higher levels. The “eff” index highlights that this is a geometrical distance in a $D$-dimensional parameter space, not the $N_\sigma$ associated with a 1D Gaussian variate, which has no direct geometric meaning for $D>1$. This interpretation of $N_\sigma^{\rm eff}$ can be generalized to the case of more than three datasets, mapping the geometry to a regular polygon (see Appendix).

This effective significance scale, $L(N_\sigma^{\rm eff})$,  
depends both on the dimension $D$ of the shared parameter space and the number of datasets $N$, and must be recomputed for each case. In particular, a similar construction can be carried out for the $Q_{\rm DM}$ statistic by mapping the problem to two posteriors with covariance $C=\frac{1}{2}\,\mathcal{I}_{D\times D}$ separated by a distance $L$.

Besides the information encoded in $\mathcal{Q}$, the dispersion tensor $\mathcal{C}$ provides additional insight into the internal structure of the tension. The determinant $\det(\mathcal{C})$ is proportional to the volume of the ellipsoid describing the distribution of tension vectors; $\det(\mathcal{C})=0$ indicates that at least one eigenvalue vanishes, meaning that the tension effectively lies in a lower-dimensional subspace. 

A second useful descriptor is the eccentricity,
\begin{equation}\label{eq: ecc}
    \mathrm{Ecc} = \sqrt{1 - \frac{\lambda_{\min}}{\lambda_{\max}}}\,,
\end{equation}
which quantifies the anisotropy of $\mathcal{C}$. When $\mathrm{Ecc}\simeq 0$, the tension vectors have comparable magnitudes, indicating that all posterior pairs exhibit similar levels of tension. In contrast, $\mathrm{Ecc}\simeq 1$ reflects large differences in vector lengths, indicating unequal contributions from the different dataset pairs to the global tension. Note that the particular case of $\mathrm{Ecc}=1$ corresponds to $\det(\mathcal{C})=0$.

\begin{figure}
    \centering
    \includegraphics[width=0.7\linewidth]{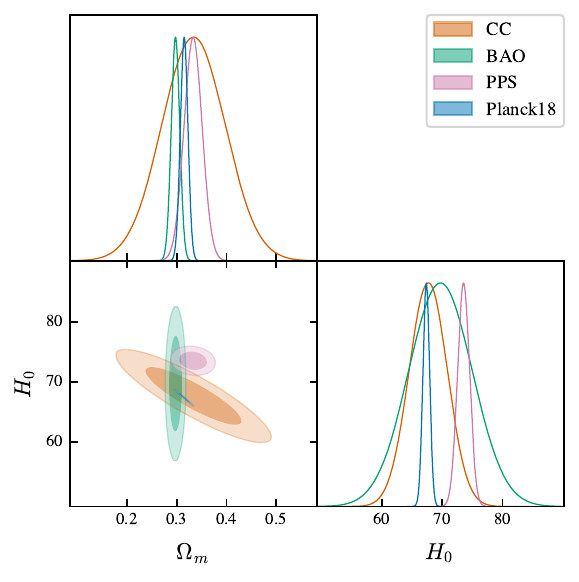}
    \caption{Example of posterior distributions under their gaussian approximation from real analyses using cosmological datasets.}
    \label{fig: posteriors}
\end{figure}

\textit{Results and Discussion ---}
We now apply our estimator to realistic examples based on posterior distributions from actual cosmological datasets (see Fig.~\ref{fig: posteriors}). Our goal is to contrast the conventional tension estimate—reported as an equivalent one-dimensional significance $N_\sigma$—with the geometric significance $N_\sigma^{\rm eff}$ derived from our method. The computational tools for these calculations are available in the public repository \texttt{MULTIMETER} \footnote{\url{www.github.com/matiasleize/multimeter}}.

Table~\ref{tab:results} shows that for low or moderate tensions (CC+Planck+BAO, CC+BAO+PPS) the two estimates are similar. However, when the disagreement is substantial (Planck+BAO+PPS, CC+Planck+PPS), the geometric significance is markedly smaller than the usual $N_\sigma$. The same pattern appears in the two–dataset case using $Q_{\rm DM}$: for instance, the tension between Planck and PPS is $3.86\sigma^{\rm eff}$ according to our geometric interpretation, whereas the standard indicator yields $5.68\sigma$. This demonstrates that the commonly quoted $N_\sigma$—constructed by analogy with a 1D Gaussian variate—can be misleading when the shared parameter space has $D>1$. We note  that the later discussion refers to the tension in the multidimensional shared parameter space, which differs from the usual ``Hubble tension'' between local and CMB-inferred values of $H_0$, where the shared parameter space is one-dimensional.
Another example of this mismatch arises from the recent release of the DESI collaboration \cite{DESIDR2} where an increase in the tension with CMB data from  $1.9 \sigma$ (DR1) to  $2.3\sigma$ (DR2) is reported while in our geometric interpretation, the change is less significant:  from $1.18\sigma^{\rm eff}$ to $1.45\sigma^{\rm eff}$. Finally, we note that the largest tensions arise in the Planck+BAO+PPS and CC+Planck+PPS combinations, as expected given that the dominant discrepancy originates from the Planck–PPS pair.

\begin{table}[]
    \centering
    \begin{tabular}{c|c|c|c}
        Datasets           & $N_\sigma$    & $L$     & $N_\sigma^{\rm eff}$\\
        \hline
        \hline
        Synthetic case 1   & $1.51$  & $2.13$  & $ 0.991$      \\
        Synthetic case 2   & $4.53$  & $3.97$  & $ 2.34$       \\
        \hline
        CC + Planck + BAO  & $0.429$ & $1.48$  & $ 0.555$      \\
        CC + BAO + PPS     & $1.78$  & $2.29$  & $ 1.11$       \\ 
        Planck + BAO + PPS & $4.1$   & $3.7$   & $ 2.14$       \\
        CC + Planck + PPS  & $4.33$  & $3.85$  & $ 2.25$       \\
        \hline
        CC + Planck        & $0.346$ & $0.794$ & $0.184$     \\
        CC + BAO           & $0.189$ & $0.569$ & $ 0.0997$     \\
        CC + PPS           & $2.57$  & $3.03$  & $ 1.64$     \\
        %Planck + BAO (D=3) & $1.15\sigma$  & $2.03$  & $ 0.581\sigma $     \\
        Planck + PPS       & $5.68$  & $6.02$  & $ 3.86$     \\
        BAO + PPS          & $1.39$  & $1.9$   & $ 0.833$     \\

    \end{tabular}
    \caption{Tension estimators for synthetic and real data sets: usual $N_\sigma$ vs our geometrical indicator $N_\sigma^{\rm eff}$ derived from the equivalent configuration distance $L$. PPS: Pantheon Plus + SH0ES \cite{Scolnic:2021amr}, CMB: Planck2018 \cite{Planckcosmo2018}, BAO: DESI DR2 \cite{DESIDR2}.}
    \label{tab:results}
\end{table}

Regarding the anisotropy in the tension, although all configurations exhibit a high level of eccentricity (see Table~\ref{tab:results_ecc}), the combination Planck+BAO+PPS yields the smallest value. This can be understood from the fact that CC is the dataset that is least in tension with the rest; consequently, when CC is excluded, the remaining datasets display more mutually consistent tension level, which in turn leads to a lower eccentricity. This behavior is clearly visible in the upper-left panel of Fig.~\ref{fig:4panel_extra}.

\begin{table}[]
    \centering
    \begin{tabular}{c|c|c|c|c|c}
        Datasets           & $\{\lambda_\alpha\}$& ${\rm Ecc}$\\
        \hline
        \hline
        Synthetic case 1   & $\{3.54,  0.\}$     & $1$         \\
        Synthetic case 2   & $\{16.8,  0.495\}$  & $0.985$     \\
        \hline
        CC + Planck + BAO  & $\{1.1,0.0456\}$    & $0.979$     \\
        CC + BAO + PPS     & $\{4.22, 0.148\}$   & $0.982$     \\ 
        Planck + BAO + PPS & $\{1.3, 12.8\}$     & $0.948$     \\
        CC + Planck + PPS  & $\{0.674, 14.7\}$   & $0.977$     \\
    \end{tabular}
    \caption{Indicators of  anisotropy in the tension for synthetic and real datasets.}
    \label{tab:results_ecc}
\end{table}

\textit{Conclusions ---} We have presented a new estimator that enables the simultaneous assessment of tensions among multiple datasets. This approach offers a novel framework for studying cosmological tensions, moving beyond pairwise comparisons to treat tension as a global property. We have also provided a geometric interpretation of the tension for the two- and three-dataset cases in a 2D shared parameter space. Unlike the conventional $N_\sigma$, which is inferred by analogy with a 1D Gaussian posterior, our $N_\sigma^{\rm eff}$ estimator  provides a direct geometric interpretation in the 2D shared parameter space for assessing dataset consistency. While these tools were developed specifically for evaluating cosmological tensions, they may find broader application in other research fields where  discrepancies between datasets emerge, like for example, the discrepancy in the value of the W boson mass in particle physics. \cite{CDF}.

%\appendix
\appendix*
\section{$\mathcal{Q}$ distribution under the null hypothesis $(H0)$} \label{sec: gamma}

Here we derive the distribution of the statistic $\mathcal{Q}$ in the case where all covariance matrices are equal. To simplify the notation, we define  
\[
D_{ij} = (\hat{C}_i + \hat{C}_j)^{-1/2} .
\]
By construction, the tension vector $\vec{r}_k$ is distributed as
\begin{equation}
\begin{split}
    \vec{r}_k &\sim \mathcal{N}\!\left(D_{ij}(\vec{\bar{\theta}}_i - \vec{\bar{\theta}}_j),\, D_{ij} (\hat{C}_i+\hat{C}_j) D_{ij}^T\right) \\
    &\sim \mathcal{N}\!\left(D_{ij}(\vec{\bar{\theta}}_i - \vec{\bar{\theta}}_j),\, I_{D\times D}\right).
\end{split}
\end{equation}

Under the null hypothesis $H_0$ defined in Eq.~\ref{eq:def-H0}, we obtain
\begin{equation}
    \vec{r}_k|_{H0} \sim \mathcal{N}(\vec{0},\, I_{D\times D}) \qquad \forall k.
\end{equation}

It is important to note that, although all $\vec{r}_k$ are identically distributed, they are not independent, since by construction they share some of the underlying random variables $\vec{\theta}_i$. The sum of the $N_p$ tension vectors therefore satisfies
\begin{equation}
    \sum_{k=1}^{N_p} \vec{r}_k|_{H0} \sim \mathcal{N}\!\left(\vec{0},\, N_p\, I_{D\times D}\right).
\end{equation}

From this, the general form of the distribution of the statistic $\mathcal{Q}$ can be expressed as a linear combination of chi-squared variables with one degree of freedom:
\begin{equation}
    \mathcal{Q}|_{H0} = \frac{1}{N_p}\sum_{k=1}^{N_p} |\vec{r}_k|^2  
    \sim \frac{1}{N_p}\sum_{i=1}^{D} a_i\, \chi^2_{1},
\end{equation}
where the coefficients $a_i$ encode the correlations among the tension vectors.  
In the particular case where all covariance matrices $\hat{C}_i$ are equal, $\mathcal{Q}|_{H0}$ reduces to
\begin{equation}
    \mathcal{Q}|_{H0} \sim \Gamma(D, 1)\, .
\end{equation}

\section{Generalization to $N$ datasets}

In order to generalize our mapping to an arbitrary number $N$ of datasets, we need to arrange the set of distributions along the vertices of a regular polygon with $N$ sides and side length $L$. To achieve this, we introduce a natural length scale $A$, defined as the longest diagonal of the polygon. It can be shown that $A$ can be expressed in terms of $N$ and $L$ as
\begin{equation} \label{eq: generalization_D}
A = \frac{\sin\left(\pi\left[\frac{N}{2}\right]/N\right)}{\sin\left(\pi/N\right)} \, L \, ,
\end{equation}
where $[\frac{N}{2}]$ denotes the integer part of $\frac{N}{2}$. Note that Eq.~\ref{eq: generalization_D} implies $A=L$ for $N=2$ and $N=3$, recovering our previous results.

For $N>3$, the quantity $N_\sigma^{\rm eff}$ represents the minimum pairwise distance between all posteriors distributed along the polygon. For instance, $N_\sigma^{\rm eff}=1$ for $N=5$ means that the five distributions are separated by at least $1\sigma^{\rm eff}$ from each other.

\begin{figure}
    \centering
    \includegraphics[width=\linewidth]{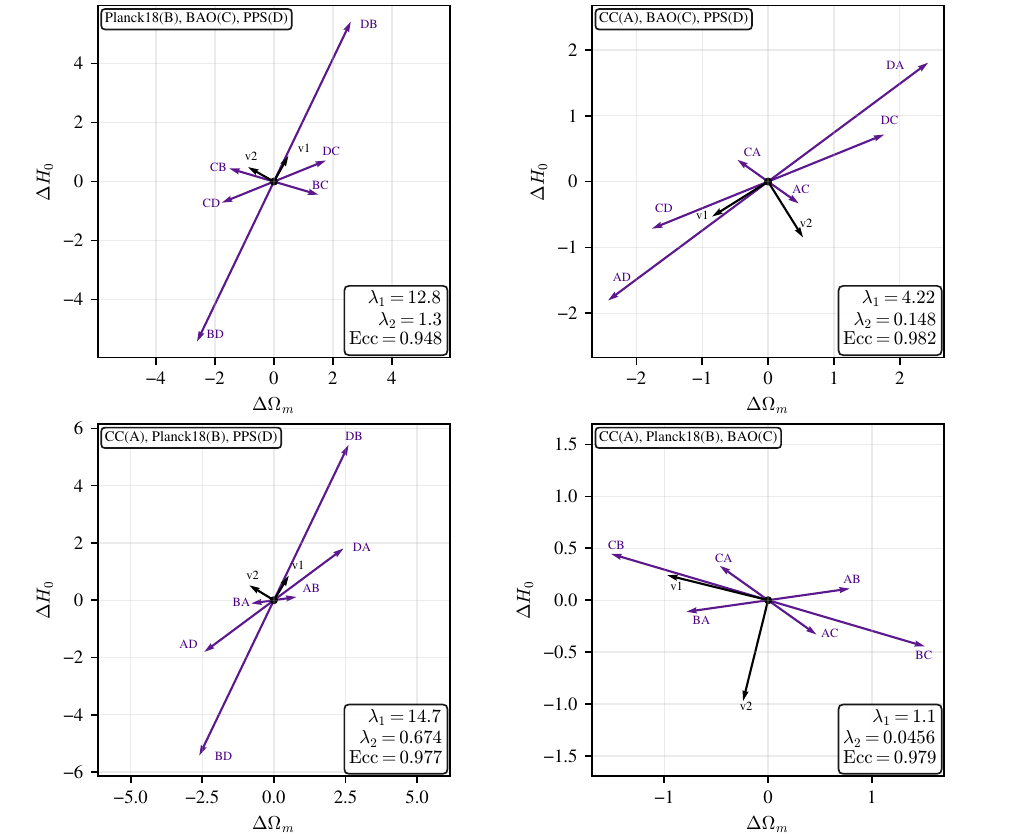}
    \caption{Sets of tension vectors for current cosmological datasets together with the corresponding eigenvalues and eccentricity.}
    \label{fig:4panel_extra}
\end{figure}

\begin{acknowledgments}
The authors thank Dario Rodrigues, Gastón F. Scialchi,  Nahuel Mirón-Granese and Esteban A. Calzetta for valuable comments and discussions. S.L. and M.L. are supported by grant PIP 11220200100729CO CONICET and grant 20020170100129BA UBACYT. C.G.S. would like to acknowledge support from the ICTP through the Associates Programme (2023-2028), and funding from CONICET (PIP-2876), and Universidad Nacional de La Plata (SG002), Argentina. 
\end{acknowledgments}

\bibliography{apssamp}
\end{document}